\documentclass[%
reprint,
superscriptaddress,
amsmath,
amssymb,aps,
]{revtex4-1}

\usepackage{graphicx}
\usepackage{dcolumn}
\usepackage{bm}

\begin{document}

\preprint{APS/123-QED}

\title{Photoneutron emission cross sections for $^{13}$C}

\author{H.~Utsunomiya}
\affiliation{Konan University, Department of Physics, 8-9-1 Okamoto, Higashinada, Japan}
\affiliation{Shanghai Advanced Research Institute, Chinese Academy of Sciences, Shanghai 201210, China}

\author{S.~Goriely}
\affiliation{Institut d'Astronomie et d'Astrophysique, Universit\'{e} Libre de Bruxelles, Campus de la Plaine, CP-226, 1050 Brussels, Belgium}


\author{M.~Kimura}
\affiliation{RIKEN Nishina Center, Saitama 351-0198, Japan}

\author{N.~Shimizu}
\affiliation{Center for Computational Sciences, University of Tsukuba, Tennodai Tsukuba 305-8577, Japan}

\author{Y.~Utsuno}
\affiliation{Advanced Science Research Center, Japan Atomic Energy Agency, Tokai, Ibaraki 319-1195, Japan}
\affiliation{Center for Nuclear Study, University of Tokyo, Hongo, Bunkyo-ku, Tokyo 113-0033, Japan}

\author{G.~M.~Tveten}
\affiliation{Department of Physics, University of Oslo, N-0316 Oslo, Norway}
\affiliation{Expert Analytics AS, N-0179 OSLO, Norway}

\author{T.~Renstr\o m}
\affiliation{Department of Physics, University of Oslo, N-0316 Oslo, Norway}
\affiliation{Expert Analytics AS, N-0179 OSLO, Norway}

\author{T.~Ari-izumi}
\affiliation{Konan University, Department of Physics, 8-9-1 Okamoto, Higashinada, Japan}

\author{S.~Miyamoto}
\affiliation{Laboratory of Advanced Science and Technology for Industry, University of Hyogo, 3-1-2 Kouto, Kamigori, Ako-gun, Hyogo 678-1205, Japan}

\date{\today}

\begin{abstract}
Photoneutron emission cross sections were measured for $^{13}$C below $2n$ threshold using quasi-monochromatic $\gamma$-ray beams produced in laser Compton-scattering at the NewSUBARU synchrotron radiation facility. The data show fine structures in the low-energy tail of the giant-dipole resonance; the integrated strength of the fine structure below 18~MeV is intermediate among the past measurements with bremsstrahlung and the positron annihilation $\gamma$ rays. We compare the photoneutron emission data with the {\sf TALYS} statistical model calculation implemented with the simple modified Lorentzian model of $E1$ and $M1$ strengths. We also compare the total photoabsorption cross sections for $^{13}$C with the shell model and antisymmetrized molecular dynamics calculations as well as the statistical model calculation. We further investigate the consistency between the present photoneutron emission and the reverse $^{12}$C(n,$\gamma$) cross sections through their corresponding astrophysical rate. 
\end{abstract}

\maketitle
\section{Introduction}
There is growing research interest in photodisintegration cross sections in the context of the origin of ultra-high energy cosmic rays (UHECRs).  The results of the Pierre Auger \cite{Aab20a,Aab20b}  and Telescope Array \cite{TA} experiments performed in the southern and northern hemispheres, respectively, are consistent with the presence of the GZK cutoff \cite{Grei66,ZatKuz66} around 10$^{20}$ eV which is inferred from the photopion production by protons and photodisintegration of nuclei in the interaction with the cosmic microwave background in the extragalactic space. However, the analyses of the distribution of atmospheric depths at which air shower develops to the maximum showed that the composition of UHECRs is highly uncertain largely due to different hadronic interaction models as well as the scarce UHECR events \cite{Auger,TA}. The effect of the extragalactic propagation of UHECRs on their composition arrived at Earth is subject to photodisintegration cross sections through the giant-dipole resonance (GDR) and models of the extragalactic background light \cite{Bati15}.  

A project called PANDORA has been launched \cite{Pandora} to investigate photonuclear data relevant to the origin of UHECRs experimentally and theoretically. The photoreaction data required are not only photoabsorption cross sections but also partial cross sections for all decay channels of GDR by photoemissions of neutrons and charged particles for nuclei less massive than the iron group (Fe-Co-Ni) nuclei. TALYS cross sections \cite{Koning12} implemented with experimental GDR parameters or theoretical dipole photon strength functions \cite{RIPL3} are widely used \cite{Bati15}. However, GDR parameters are not well elucidated experimentally for the less-massive nuclei and theoretical models often fail to predict the correct photon strength function properties. 

Photonuclear data for $^{13}$C compiled in the experimental nuclear data library EXFOR \cite{EXFOR} are photoneutron \cite{Cook57,Koch76,Jury79,Wood79} and photoproton \cite{Cook57,Denis64,Zuban83} emission cross sections. The total photoabsorption cross section is also estimated from these photoneutron and photoproton cross sections \cite{Ishk02}.  
In this paper, we report photoneutron emission cross sections for $^{13}$C measured with quasi-monochromatic $\gamma$-ray beams produced in the laser Compton scattering.  We discuss photonuclear data for $^{13}$C in comparison with the statistical model, shell model, and antisymmetrized molecular dynamics  (AMD) calculations.  We also discuss the stellar photodissociation rate of $^{13}$C which is linked to the Maxwellian-averaged $^{12}$C$(n,\gamma)$ cross section by the detailed balance theorem.   

\section{Experimental procedure} 
\label{sec_exp}
Quasi-monochromatic $\gamma$-ray beams were produced in the laser Compton scattering (LCS) of 1064 nm photons from the Inazuma Nd:YVO$_4$ laser with relativistic electrons in the NewSUBARU storage ring.  A linear accelerator was used to inject electrons at a fixed energy of 974 MeV into the storage ring. The injected electrons were either decelerated to 743 MeV or accelerated to 1149 MeV to produce LCS $\gamma$-ray beams in the energy range of 10.01 - 23.48 MeV below $2n$ threshold at 23.67 MeV. The electron beam energy has been calibrated with an accuracy on the order of 10$^{-5}$ \cite{Utsu14,Shima}.  The reproducibility of the electron energy is assured by an automated control of the electron beam-optics parameters~\cite{Utsu14}.  LCS $\gamma$-ray beams passed through two 10 cm-long Pb collimators with 3 mm and 2 mm apertures and were delivered to the experimental hutch GACKO (gamma collaboration hutch of Konan University).  

\begin{figure}
\includegraphics[scale=0.5]{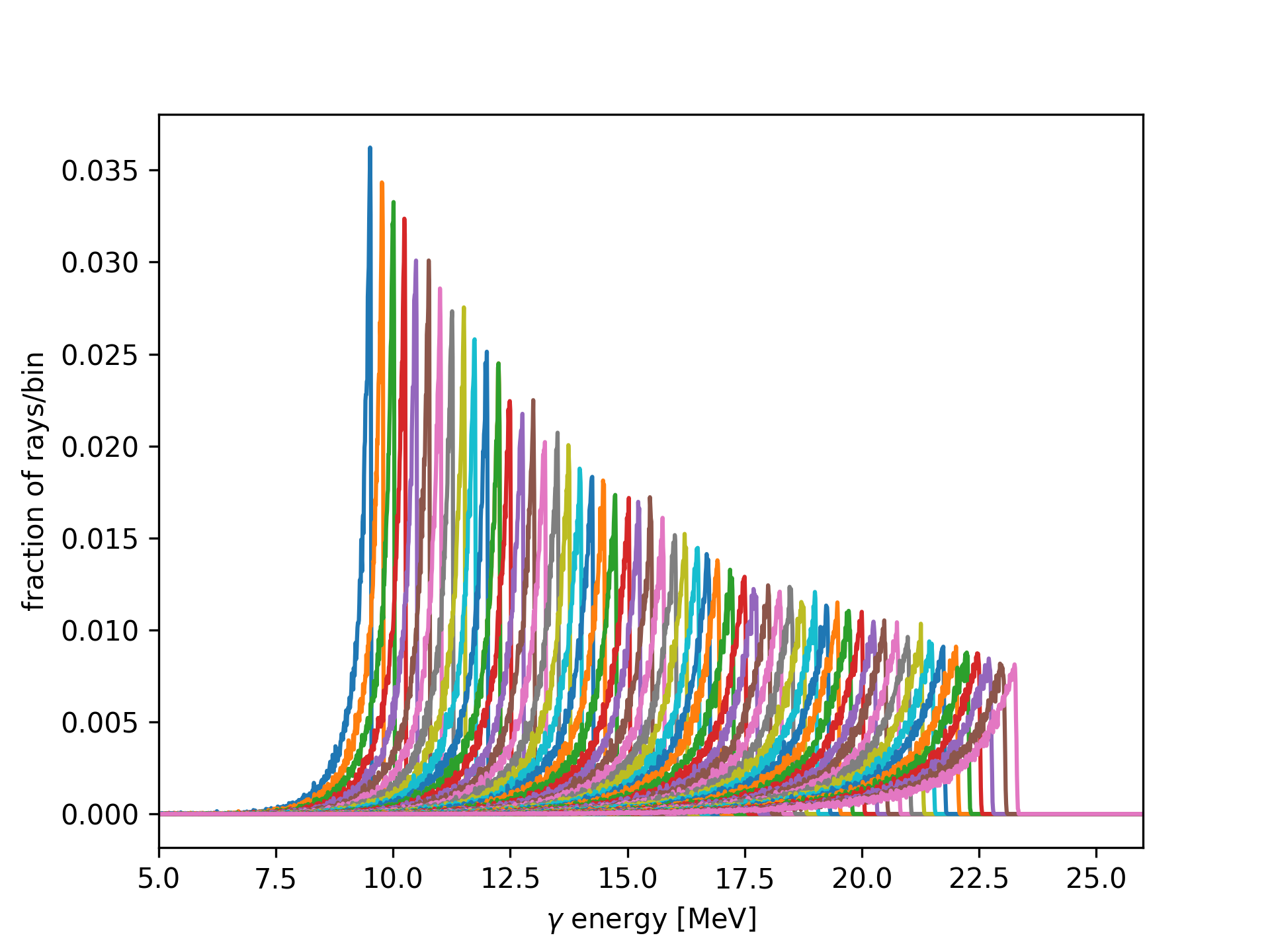}
\caption{(Color online)  The incident $\gamma$-ray beams. The profiles have been normalized and the energy bins are 10 keV.}  
\label{fig:incident_lcs}
\end{figure}

Response functions of the LCS $\gamma$-ray beams were measured with a $3.5^{\prime\prime}\times 4.0^{\prime\prime}$ LaBr$_3$(Ce) (LaBr$_3$) detector. The energy profiles of the LCS $\gamma$-ray beams were determined by best reproducing the response functions in Monte Carlo simulations with a GEANT4 code~\cite{Ioana_thesis, geant4ref} that incorporates the kinematics of the LCS process, including the electron beam emittance \cite{Utsu15,Fili14}. The incident LCS $\gamma$-ray beams are shown in Fig. \ref{fig:incident_lcs}. The $\gamma$-ray flux was monitored with a $8^{\prime\prime} \times 12^{\prime\prime}$ NaI(Tl) (NaI) detector. The number of LCS $\gamma$ rays were determined with the pile-up or Poisson-fitting method for pulsed $\gamma$-ray beams~\cite{Utsu18a,Kond11}.  

The $^{13}$C target was in amorphous form with the isotopic enrichment of 99 \% and chemical purity of 97 \%. The target material was packed into two aluminum cylindrical containers of 8 mm inner diameter with the entrance and exit windows of 25.4 $\mu$ m Kapton foils. The total areal density was 2940 mg/cm$^2$.    

Photoneutrons were measured with the high-efficiency $4\pi$ detector consisting of 10 bars $^3\rm{He}$ proportional counters of 25 mm diameter and 45 cm length that were embedded in a 36 $\times$ 36 $\times$ 50 cm$^3$ polyethylene neutron moderator in three concentric rings of 4, 8, and 8  $^3\rm{He}$ counters at the distance of 3.8, 7.0, 10.0 cm , respectively, from the axis of the LCS $\gamma$-ray beam~\cite{neutrondet}. The efficiency of the neutron detector varies with the neutron kinetic energy. The efficiency was calibrated with a $^{252}$Cf source with the emission rate of 2.27 $\times$ 10$^4$ s$^{-1}$ with 2.2\% uncertainty \cite{Nyhu15} and the energy dependence was determined by MCNP Monte Carlo simulations \cite{MCNP}. The ring ratio technique, originally developed by Berman {\it et al.}~\cite{Berman_ring_ratio}, was used to determine the average energy of neutrons emitted in the $(\gamma,n)$ reaction. The average energy was used to determine the neutron detection efficiency.  During the neutron measurement, the laser was periodically turned on for 80 ms and off for 20 ms in every 100 ms, to measure background neutrons.  A blank target container was used to measure neutrons from the Kapton foils in every photoneutron measurement. The contribution from the Kapton foils turned out to be negligible.  

\section{Unfolding method}

The neutron yield cross section experimentally determined by the number of incident LCS $\gamma$ rays ($N_{\gamma}$), the areal density of target nuclei ($N_t$), and the number of neutrons detected ($N_n$) represents a quantity that is expressed by folding the photoneutron emission cross section $\sigma(E_{\gamma})$ with the energy distribution of the LCS $\gamma$-ray beam \cite{Renst18}).  

\begin{equation}
\int_{S_n}^{E_{\rm max}}D^{E_{\rm max}}(E_{\gamma})\sigma(E_{\gamma})dE_{\gamma}=\frac{N_n}{N_tN_{\gamma}\xi\epsilon_n g}=\sigma_{exp}^{E_{max}}.
\label{eq:cross1}
\end{equation}
Here, $D^{E_{\rm max}}$ is the normalized energy distribution of the $\gamma$-ray beam shown in Fig.~\ref{fig:incident_lcs}, $\int_{S_n}^{E_{\rm max}} D^{E_{\rm max}}dE_{\gamma}= 1$. 
The quantity $\epsilon_n$ represents the neutron detection efficiency and $\xi=(1-e^{-\mu t})/(\mu t)$ gives a correction factor for self-attenuation in the target. The factor $g$ represents the fraction of the $\gamma$ flux above $S_n$. 

For $\gamma$-ray energy distributions with the $\delta$ function, the integral equation (Eq.~(\ref{eq:cross1})) is obviously unfolded to a monochromatic cross section.  Experimentally, cross sections are determined in the monochromatic approximation which we plot at the maximum energy of the LCS $\gamma$-ray beam $E_{max}$ as the experimental cross section, $\sigma_{exp}^{E_{max}}$.   

\begin{figure}
\includegraphics[scale=0.45]{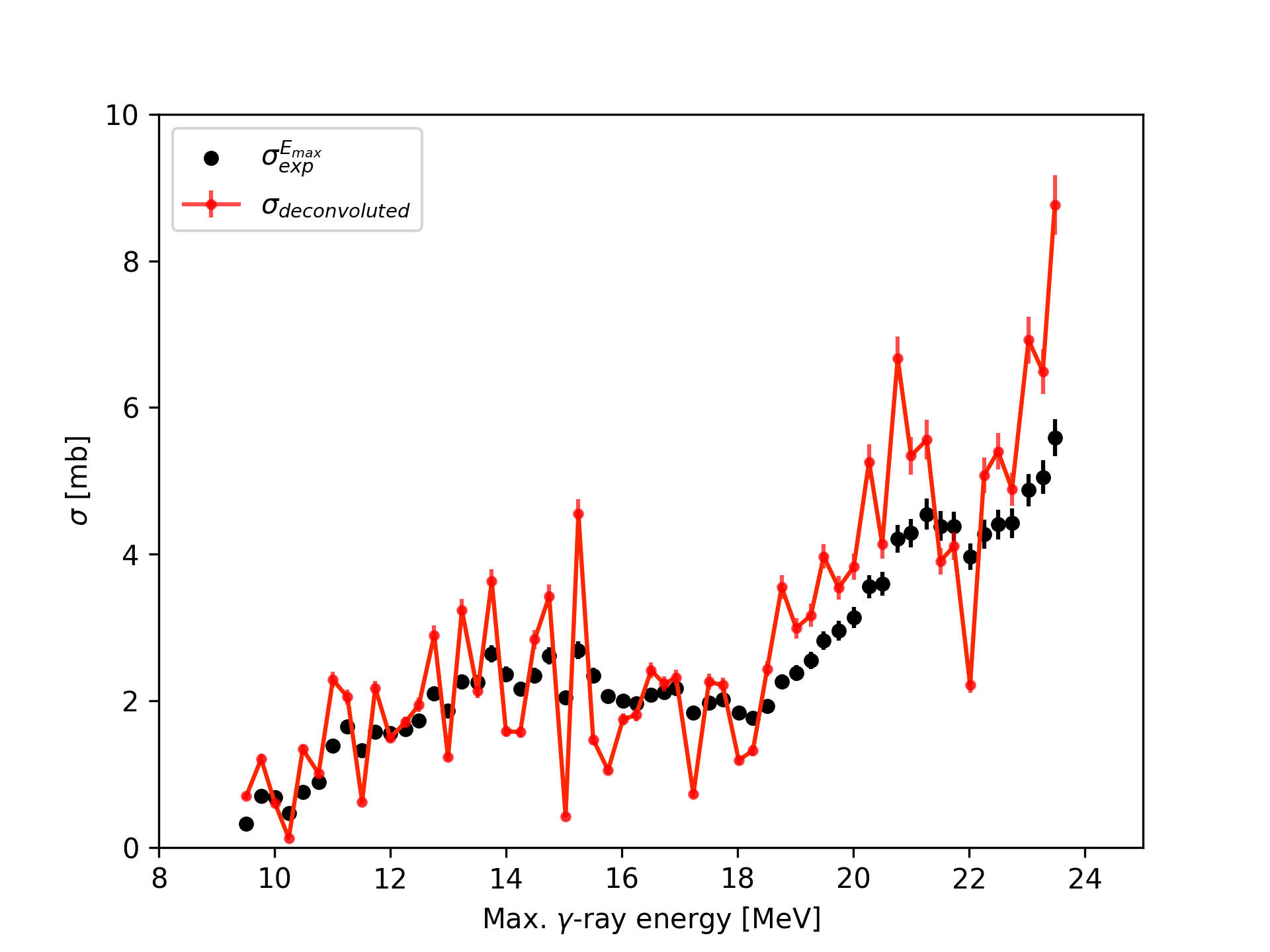}
\caption{(Color online) The monochromatic cross section measured per incident LCS $\gamma$-ray beam, $\sigma_{exp}^{E_{max}}$, and the unfolded cross section, $\sigma_{deconvoluted}$.}  
\label{fig:mono_unfold}
\end{figure}

Photoneutron emission cross sections $\sigma(E_{\gamma})$ in Eq.~(\ref{eq:cross1}) were unfolded at $E_{max}$ with the deconvolution method \cite{Renst18}. The method was routinely applied to the data of Ni \cite{Utsu18b}, Tl \cite{Utsu19a}, and Ba \cite{Utsu19b} isotopes.  
The maximum energy of the incident LCS $\gamma$-ray beams in the current experiment was made in rather small increments and therefore the interpolation was done with a first order polynomial in numerical iterations of solving a set of linear equations (Eq.~(3) of Ref.~\cite{Renst18}) toward a convergence of the reduced $\chi^2$ to unity.  This ensures that no spurious fluctuations are caused by the choice of parameters for the cubic spline. 
Furthermore, we have ensured that the unfolding did not introduce spurious structures that are due to a statistical coincidence by repeating the unfolding 1000 times where the experimental cross section measured per incident LCS $\gamma$-ray beam was randomly sampled from a Gaussian distribution where the standard deviation parameter was set to be the total experimental uncertainty for the measured $\sigma_{exp}^{E_{max}}$ at $E_{max}$. The fine structure has been resolved in the repeated unfolding procedure. Fig. \ref{fig:mono_unfold} shows unfolded cross sections with the total uncertainty in comparison with the monochromatic cross sections. The total uncertainty was estimated by the error propagation of the 1$\sigma$ uncertainty for the monochromatic cross section through the unfolding \cite{Utsu18b,Utsu19a,Utsu19b}.  

\section{Result}
\label{sec_data}
\subsection{Photoneutron emission cross section}
Previously photoneutron emission cross sections were measured for $^{13}$C by neutron counting in irradiations with bremsstrahlung \cite{Cook57,Koch76} and positron annihilation $\gamma$-ray beams \cite{Jury79}. Following the nomenclature of the IAEA photonuclear data library \cite{Kawa20}, these cross sections referred to as the inclusive one-neutron emission cross section are expressed as $\sigma(\gamma, 1nX)$ which reads  $\sigma(\gamma, 1nX)$=$\sigma(\gamma, n)$ + $\sigma(\gamma, np)$ + $\sigma(\gamma, n\alpha)$ + ... , where $X$ stands for anything except for the detected one neutron. Bremsstrahlung radiations were also used to determine 
$(\gamma,p)$ cross sections by detecting the $\beta$ activity from $^{12}$B \cite{Cook57,Denis64,Zuban83}. Figure 1 shows the present $(\gamma,1nX)$ cross sections for $^{13}$C in comparison with the previous data. 
The present data show fine structures in the low-energy tail of giant dipole resonance. The integrated strength of the fine structure below 18~MeV is lower than the positron annihilation data \cite{Jury79} and bremsstrahlung data of Cook et al. \cite{Cook57} and higher than the bremsstrahlung data of Koch et al. \cite{Koch76}.   Above 18~MeV, the present $(\gamma,1nX)$ cross section satisfactorily agrees with the positron annihilation data  \cite{Jury79}.  

\begin{figure}
\vspace{3cm}
\includegraphics[bb = 100 170 470 425, scale=0.5]{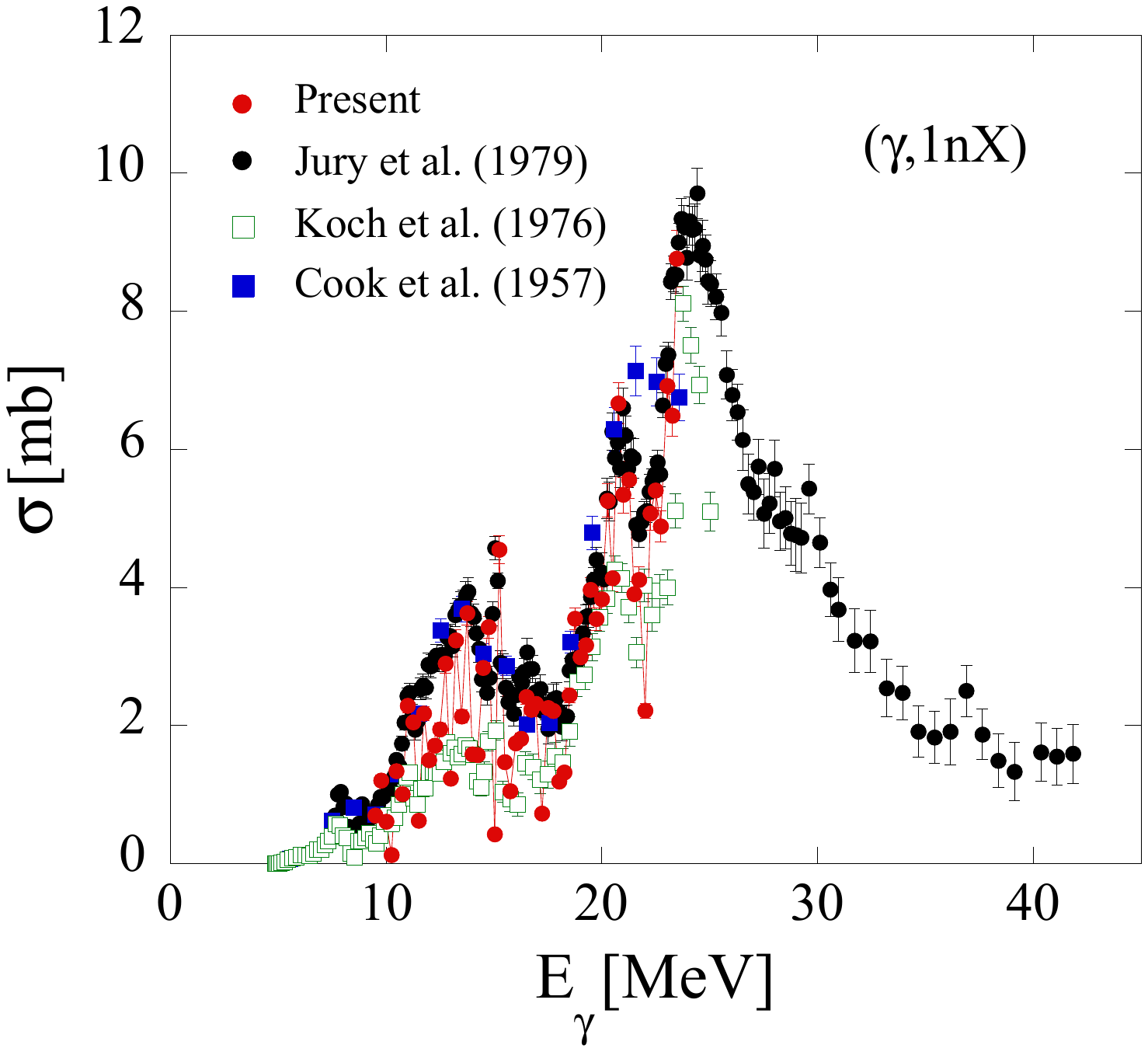}
\caption{(Color online)  Photoneutron emission cross sections, $\sigma(\gamma, 1nX)$, for $^{13}$C (red filled circles). The bremsstrahlung data of \cite{Cook57} and \cite{Koch76} are shown by blue filled squares and open green squares, respectively. The data obtained with the positron annihilation $\gamma$ rays \cite{Jury79} are shown by black filled circles. }  
\label{fig:gn}
\end{figure}

\subsection{Photoabsorption cross section}

\begin{figure}
\begin{center}
      \includegraphics[scale=0.33] {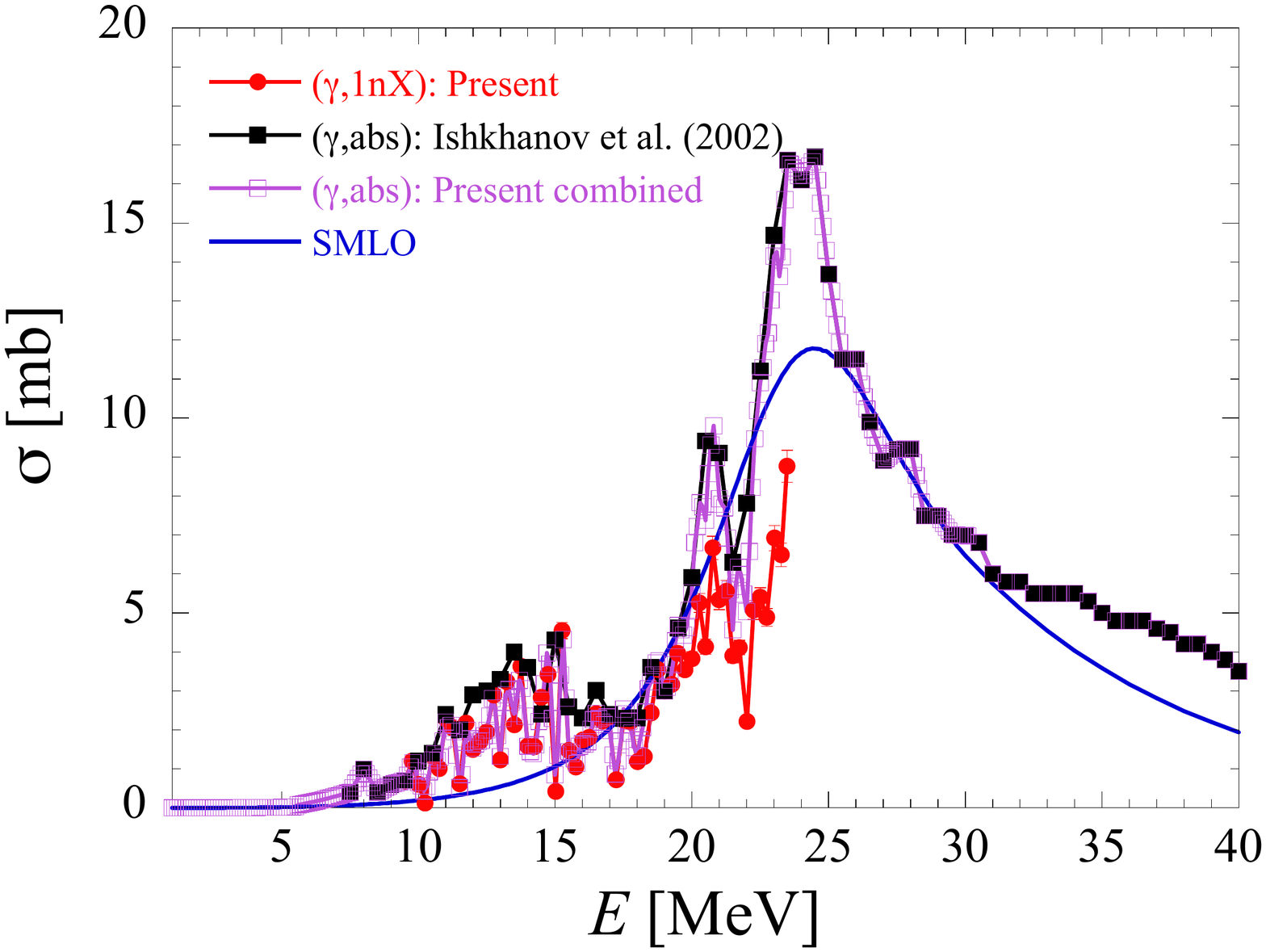}
\caption{\label{fig_13Cgabs} (Color online) Comparison of the present total photoabsorption cross section with the total photoabsorption of  \cite{Ishk02}.  Photoneutron emission cross sections \cite{Cook57,Jury79} used in the estimate of the photoabsorption are also shown.   
The blue solid line corresponds to the SMLO prediction \cite{Plujko18}.}
\end{center}
\end{figure}

To estimate $^{13}$C total photoabsorption cross section, different missing contributions have been added to the present photoneutron emission cross section. These include the photoproton emission as measured in \cite{Zuban83} as well as the $(\gamma,2n)$ + $(\gamma,2np)$ measured in \cite{Jury79}. The photoneutron cross section has in addition been supplemented at low energies with the measurement of \cite{Cook57} at E $<$ 7.5 MeV and \cite{Jury79} at energies E $<$ 9.5 MeV. All these contributions have already been taken into account by the total photoabsorption cross section estimated in \cite{Ishk02}. 
For this reason, we have used this total photoabsorption cross section for which we replaced the photoneutron cross section \cite{Jury79} with our newly measured cross section.
The final combined cross section is shown in Fig.~\ref{fig_13Cgabs} in comparison with the one of \cite{Ishk02}. 

\section{Discussion}
\label{sec_disc}

\subsection{Comparison with statistical model predictions}     

\begin{figure}
\begin{center}
      \includegraphics[scale=0.33] {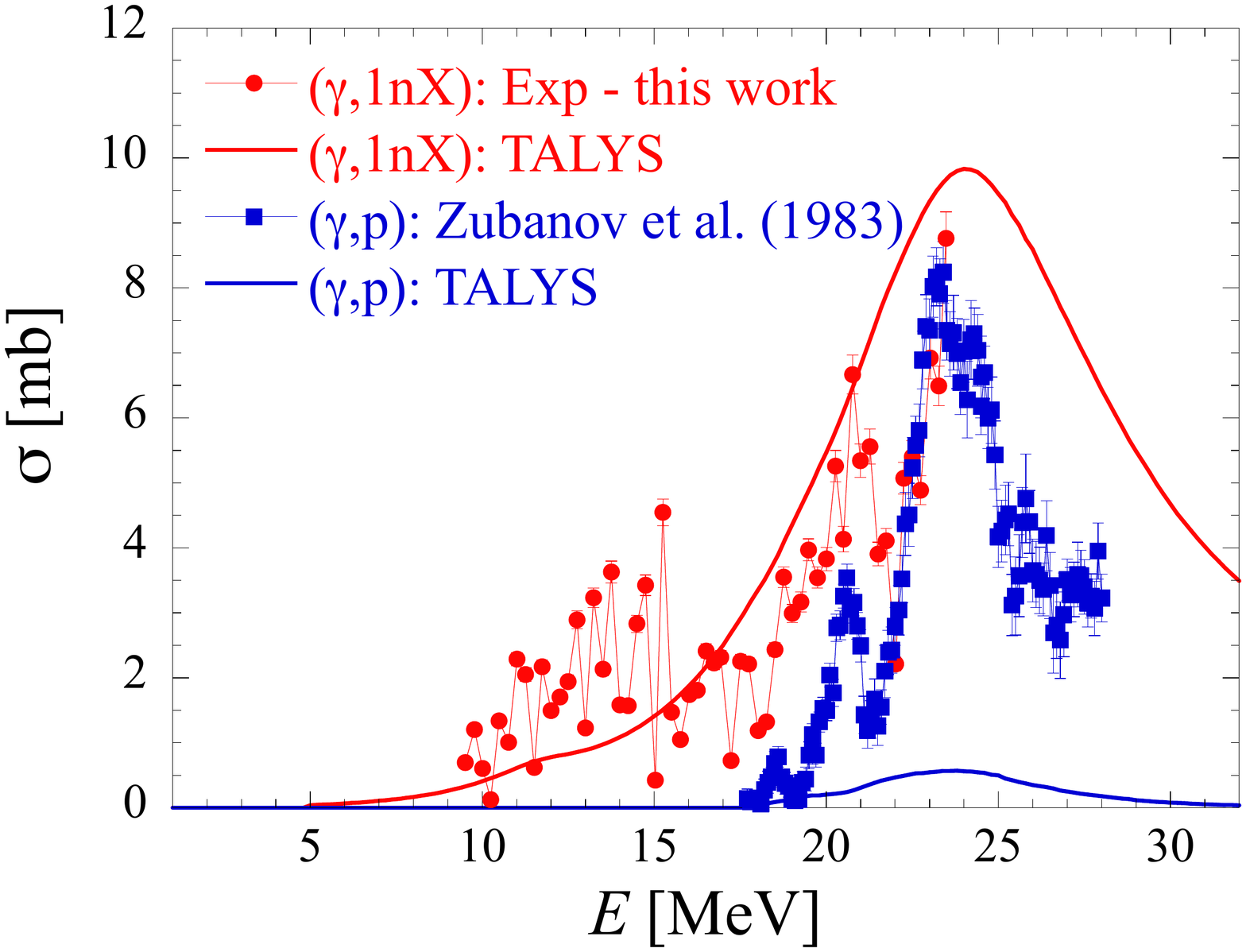}
\caption{\label{fig_13Cgn} (Color online) Comparison of the present experimental photoneutron emission cross section (red dots) and the photoproton emission cross section of \cite{Zuban83} with {\sf TALYS}  calculations based on the SMLO E1 and M1 strength functions.}
\end{center}
\end{figure}

Based on the statistical model of Hauser-Feshbach, the {\sf TALYS} code is known to be successful to estimate the cross section for medium-mass and heavy target nuclei. However, such a statistical approach is not well suited for the description of the reaction mechanisms taking place with light species such as  $^{13}$C. {\sf TALYS} may still be applied and tested for this specific photoreaction. To do so, different models of the dipole strength function are available in {\sf TALYS}, but for light species these models are not particularly accurate if not guided and adjusted directly on experimental data. Microscopic models of the QRPA-type (see e.g. Ref.~\cite{Goriely18b}) are usually not extended down to elements as light as C. In contrast the Simple Modified Lorentzian model (SMLO) for both the E1 and M1 strength functions \cite{Plujko18,Goriely18a} can be applied. 
Note that the TALYS code accounts, close to the $N=Z$ line,  for isospin forbidden transitions both in the single and multiple particle emission channels through a phenomenological correction reflecting the hindrance of dipole emission in self-conjugate nuclei, as introduced in Ref.~\cite{Holmes76}.
While the SMLO model roughly describes the $(\gamma,1nX)$ channel, the $(\gamma,p)$ cross section is underestimated by about 1 order of magnitude.
The phenomenological corrections for the isospin forbidden transitions, as explained above, affect the partial cross section predictions shown in Fig.~\ref{fig_13Cgn}, but not in a way capable of explaining the discrepancies in the $(\gamma,p)$ channel between experiments and calculations.
This shows how important it remains to measure the cross sections for such light species. The total photoabsorption cross section is similarly underestimated by the SMLO prediction as shown in Fig.~\ref{fig_13Cgabs}.  

\subsection{shell model calculation}
\begin{figure}
\begin{center}
\vspace{-1cm}
      \includegraphics[scale=0.45] {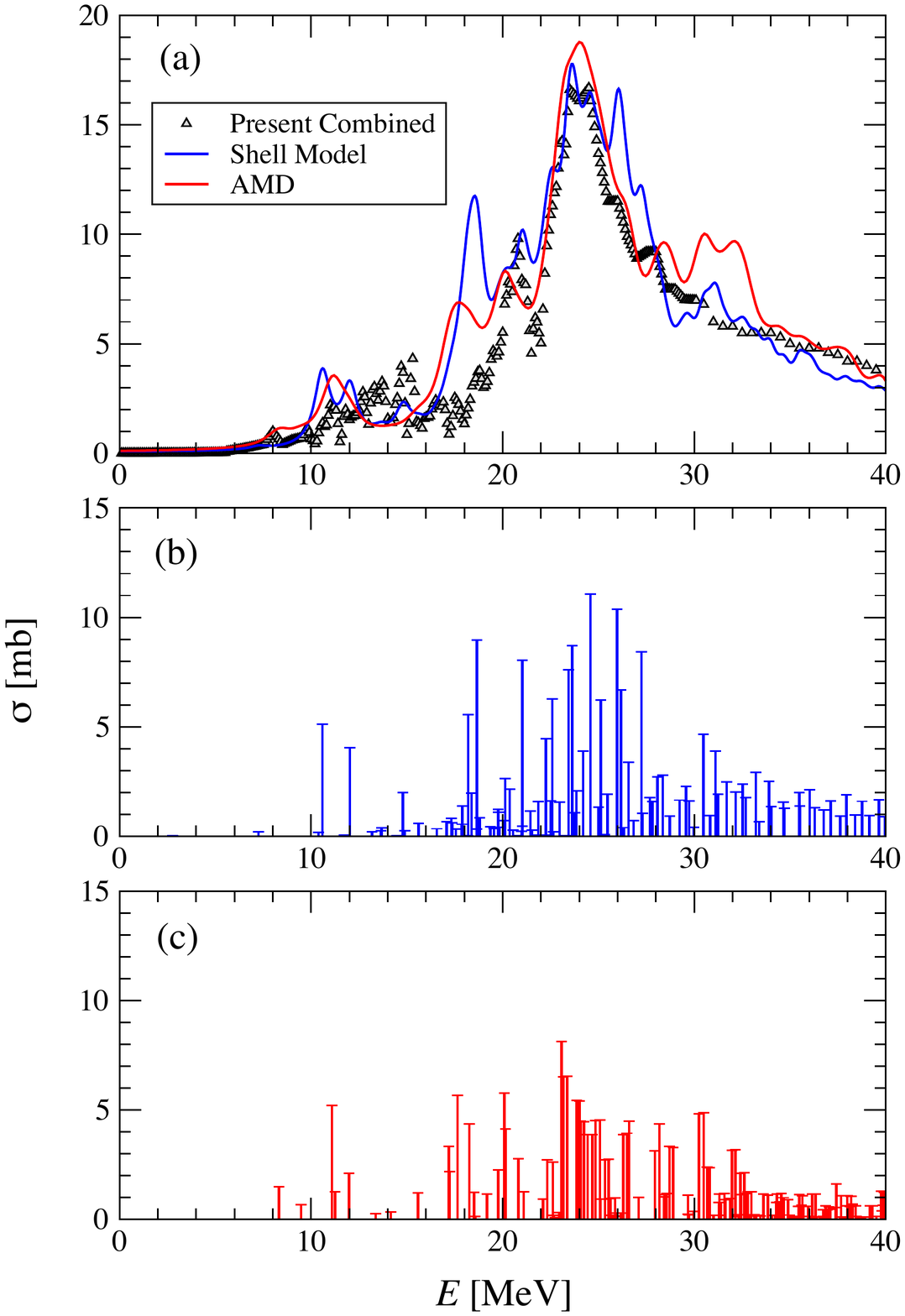}
\caption{\label{fig:C13_sm_amd} (Color online) (a) Results of the shell-model calculation (blue) and AMD calculation (red) after smoothing with $2\gamma =1$~MeV in comparison with the total photoabsorption cross section recommended. An energy shift by - 1.5 MeV is introduced to the AMD calculation. (b) Distribution of E1 excitations in the shell-model calculation with 300 Lanczos iterations. (c) Distribution of E1 excitations in the AMD calculation with an energy shift by -1.5 MeV. }
\vspace{-1cm}
\end{center}
\end{figure}

Shell-model calculations were carried out with the WBT interaction 
\cite{wbt} in the $N_{\textrm{HO}}=0$ to 3 valence shell, where 
$N_{\textrm{HO}}$ stands for the harmonic-oscillator quantum number 
of the single-particle wave functions. 
This interaction is among the most frequently used ones for the $p$-shell 
region, with two-body matrix elements fine-tuned so that experimental 
binding energies are well reproduced. 
While this interaction was originally used in the pure $0\hbar\omega$, 
$1\hbar\omega$, and $2\hbar\omega$ model spaces within this valence shell, 
here we calculated the ground state and the $E1$-excited states 
in the $(0+2)\hbar\omega$ and $(1+3)\hbar\omega$ spaces measured 
from the lowest $\hbar\omega$ state, respectively. 
To efficiently obtain the distribution of $E1$ excitations, we 
employed the Lanczos strength function method \cite{Caurier05} 
with 300 Lanczos iterations. 
We used the standard charges, $(e_p, e_n)=((N/A)e, -(Z/A)e)$, for 
calculating $E1$ strengths. 
The shell-model calculations were performed with the KSHELL code 
\cite{Shimizu19}.
The cross sections of photonuclear reactions were obtained with
\begin{equation}
 \sigma(E_{\gamma}) = \frac{16\pi^3}{9\hbar c} \sum_{\nu} 
(E_{\nu}-E_{g.s.})
B(E1;g.s. \to \nu)f(E_{\gamma}; E_{\nu}, \gamma) ,\label{eq:sm1}
\end{equation}
where $f(E_{\gamma};E_{\nu},\gamma)$ is the Lorentzian function defined by 
\begin{equation}
 f(x; x_0, \gamma) = \frac{1}{\pi}\frac{\gamma}{(x-x_0)^2+\gamma^2} .\label{eq:sm2}
\end{equation}
This function was introduced, as usual, to obtain smooth distributions 
from the discrete shell-model spectra. 
We took the FWHM of $\Gamma = 2\gamma =1$~MeV. 
Results of the shell-model calculation for $E1$ excitations are shown in Fig.~\ref{fig:C13_sm_amd}(a) after the smoothing with $2\gamma =1$~MeV and in Fig.~\ref{fig:C13_sm_amd}(b) before the smoothing.   

\subsection{antisymmetrized molecular dynamics calculation}
AMD calculations were carried out with the Gogny D1S density functional
\cite{Gogny} which has been already applied to the study of the electric dipole response of the
neutron-rich nuclei such as $^{26}{\rm Ne}$~\cite{Ne26}. The ground state is calculated with the
generator coordinate method using the quadrupole deformation as the generator coordinate.

To describe the $E1$ responce of $^{13}{\rm C}$, we employed the shifted-basis
method\cite{Ne26} which superposes the spatially shifted single-particle wave functions. The
obtained discrete energy distribution of the $E1$ strength is smeared with the Lorentzian function.
The cross sections of photonuclear reactions were calculated by using the same equation with the shell model calculations, Eqs.~(\ref{eq:sm1}) and~(\ref{eq:sm2}) with the same parameter of 2$\gamma =
1.0$ MeV. 
As is reported in the QRPA calculation \cite{Peru}, AMD calculations with the the Gogny D1S interaction also need a phenomenological correction of a shift of the E1 strength to lower energies due to the contribution beyond the one-particle-one-hole excitations. 
Results are shown in Fig.~\ref{fig:C13_sm_amd}(a) after the smoothing with $2\gamma =1$~MeV and in Fig.~\ref{fig:C13_sm_amd}(c) before the smoothing.  An energy shift by -1.5 MeV is introduced in Figs.~\ref{fig:C13_sm_amd}(a) and \ref{fig:C13_sm_amd}(c) to reproduce the peak energy of the total photoabsorption cross section.

\subsection{Comparison with inverse (n,$\gamma$) rates}     

Assuming none of the $^{13}$C excited states can be thermally populated, it is straightforward to estimate the total $^{13}$C($\gamma$,n)$^{12}$C stellar photodissociation rate from the photoneutron cross section $\sigma_{(\gamma,n)}$ on the basis of the black-body Planck distribution at a given temperature $n_\gamma(E,T)$, {\it i.e.}
\begin{eqnarray}
\lambda_{(\gamma,n)}(T)&=&\int_0^\infty c n_\gamma(E,T) \sigma_{(\gamma,n)}(E) dE \label{eq_lam} \\
&=&\frac{8 \pi}{h^3c^2} \int_0^\infty \frac{E^2}{\exp(E/kT)-1}\sigma_{(\gamma,n)}(E) dE ~.\nonumber
\end{eqnarray}
where $c$ is the speed of light and $h$ the Planck constant.
To estimate the rate $\lambda_{(\gamma,n)}$, the present photoneutron cross section $\sigma_{(\gamma,n)}$ has been supplemented at low energies with the measurement of Ref.~\cite{Cook57} at $E<7.5$~MeV and Ref.~\cite{Jury79} at energies $E < 9.5$~MeV.

\begin{figure}
\begin{center}
      \includegraphics[scale=0.4] {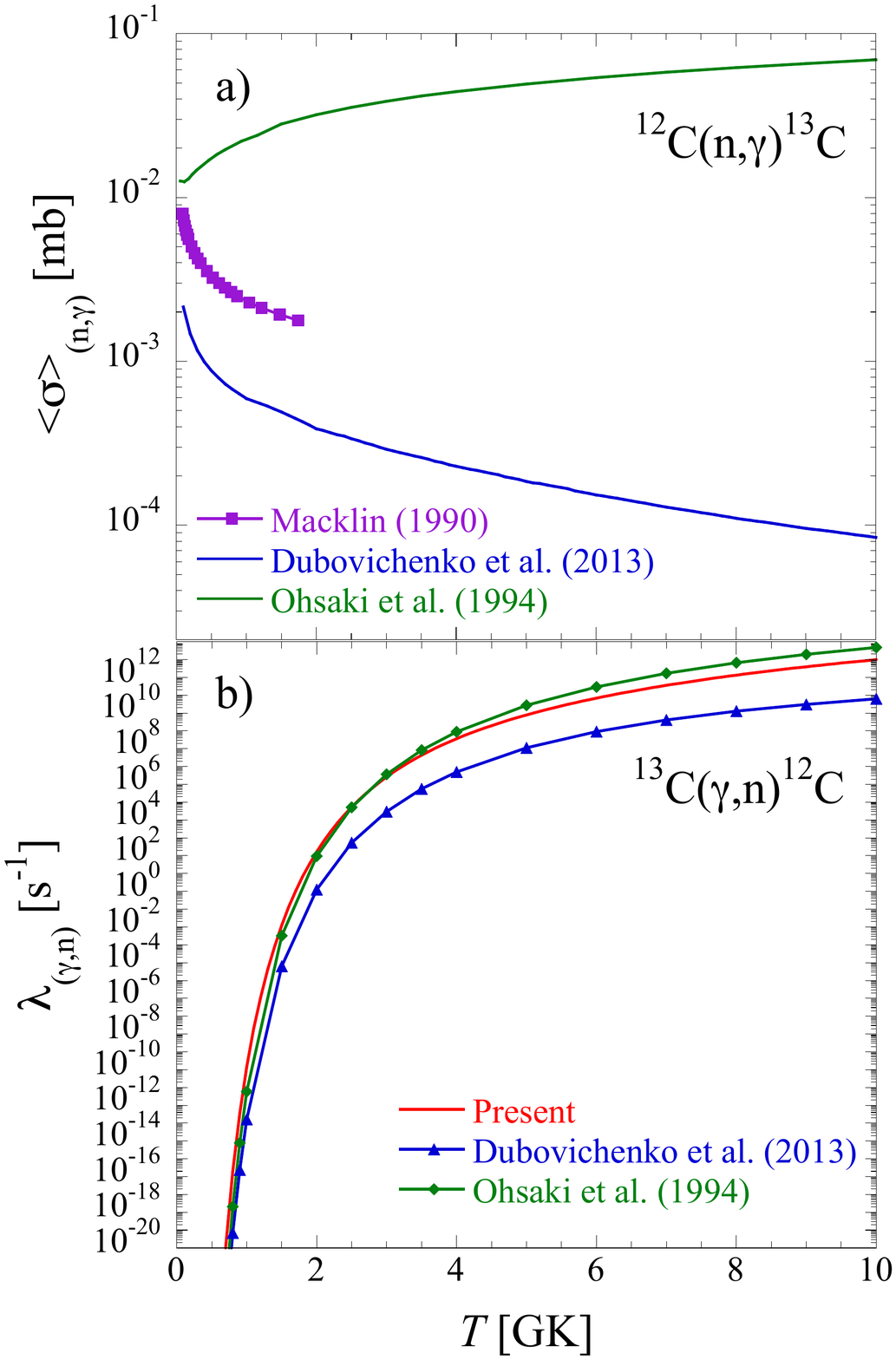}
\caption{\label{fig_ngn} (Color online) a) Experimental $^{12}$C(n,$\gamma$)$^{13}$C Maxwellian-averaged cross sections \cite{Macklin90,Ohsaki94,Dubovichenko13} as a function of the temperature $T$. b) Comparison between the $^{13}$C($\gamma$,n)$^{12}$C rates deduced from the inverse experimental cross sections of Refs. \cite{Ohsaki94,Dubovichenko13} shown in a) and the Planck-averaged rate estimated from the present experimental photoneutron emission of $^{13}$C. }
\end{center}
\end{figure}

Alternatively, making use of the detailed balance theorem, it is possible to estimate the Planck-averaged photoneutron emission rate of $^{13}$C from the inverse $^{12}$C radiative neutron capture rate  $\langle\sigma v\rangle$, {\it i.e.}
\begin{equation}
\lambda_{(\gamma,n)}(T) =  \frac{G_{^{12}C}(T)}{G_{^{13}C}(T)}
 \Bigl( \frac{m kT}{2 \pi \hbar^2} \Bigr)^{3/2} \times  \langle\sigma v\rangle_{(n,\gamma)}
~ {\rm e}^{-S_n /kT},
\label{eq_inverserate}
\end{equation}
where $G(T)$ is the temperature-dependent partition function, $S_n=4.946$~MeV the neutron separation energy, $m$ the reduced mass and $k$ the Boltzmann constant.
The $^{12}$C(n,$\gamma$)$^{13}$C Maxwellian-averaged cross section has been measured \cite{Macklin90,Ohsaki94} and studied within the framework of the potential cluster model \cite{Dubovichenko13}, as shown in Fig.~\ref{fig_ngn}. Major discrepancies exist between these different determinations, even regarding the energy dependence. These cross sections have been used to estimate the Maxwellian-averaged rate and, from it, the inverse $^{13}$C($\gamma$,n)$^{12}$C rate, as shown in Fig.~\ref{fig_ngn}b that can be compared with the Planck-averaged rate estimated (Eq.~\ref{eq_lam}) from the present experimental photoneutron emission of $^{13}$C. 
The contribution of the new data above 9.5~MeV remains small compared with those in the energy range just above the neutron separation energy of 4.9~MeV. However, the new data are found still to affect the photorate shown in Fig. 5 by a factor of 2 to 6. It will be seen that the present rate, including the SMLO extrapolation at low energies, tends to favor the $^{12}$C(n,$\gamma$)$^{13}$C cross section measured in Ref.~\cite{Ohsaki94}.

\section{Conclusion}
\label{sec_conc}
The present measurement for $^{13}$C performed below $2n$ threshold with LCS $\gamma$-rays has shown the fine structure of $(\gamma,1nX)$ cross sections in the low-energy tail of GDR. The integrated strength of the fine structure below 18~MeV is intermediate among the past bremsstrahlung \cite{Cook57,Koch76}
and positron annihilation \cite{Jury79} data. The discussion of the stellar photodissociation rate through the detailed balance theorem showed that the present photoneutron emission cross section for $^{13}$C is consistent with the $^{12}$C$(n,\gamma)$ cross section reported in Ref. \cite{Ohsaki94}. 

A recommended total photoabsorption cross section for $^{13}$C was reconstructed based on the present data supplemented with the missing contributions from photoneutron \cite{Cook57,Jury79} and photoproton \cite{Zuban83} emission cross sections. Hauser-Feshbach statistical model calculations were performed with the {\sf TALYS} code using the SMLO model of E1 and M1 strengths. 
While the statistical model roughly reproduced the experimental $(\gamma,1nX)$ cross section, the model underestimated the reconstructed total photoabsorption cross section by 40\% and the $(\gamma,p)$ cross section by an order of magnitude. 
Although {\sf TALYS} cross sections are widely used in the simulation of the extragalactic propagation of UHECRs \cite{Bati15}, the statistical nature of photodissociation cross sections needs to be further investigated for nuclei with $A$ in a comparable mass region. The shell-model and AMD calculations with a phenomenological energy shift in the latter case reasonably reproduce the total photoabsorption cross section for $^{13}$C, showing advantage over QRPA and statistical model calculations in the predictability for light-mass nuclei relevant to the nuclear origin of UHECRs.

\section{Acknowledgments}
The authors are grateful to H. Ohgaki of the Institute of Advanced Energy, Kyoto University for making a large volume LaBr$_3$(Ce) detector available for the experiment. The authors also wish to thank S. Katayama, B.V. Kheswa and D. Takenaka for contributing with taking shifts.   
H.U. acknowledges the support from the Chinese Academy of Sciences President's International Fellowship Initiative, Grant No. 2021VMA0025.  S.G. acknowledges the support from the F.R.S.-FNRS. G.M.T. acknowledges funding from the Research Council of Norway, Project Grant Nos. 262952. Y.U. acknowledges the support from JSPS KAKENHI Grant No. 20K03981. This work was supported by the IAEA and performed in line with the IAEA CRP on "Updating the Photonuclear data Library and generating a Reference Database for Photon Strength Functions'' (F41032).

\bibliographystyle{apsrev4-1}
\bibliography{newsubarubibfile}

\end{document}